\documentclass{article}

\usepackage{amsmath,amsfonts,amssymb}

\newcommand{\BR}{{\mathbb R}}

\newcommand{\tr}{\hbox{ Tr}}

\begin{document}
\title{A strong coupling expansion for $N=4$ SYM theory and other SCFT's.}
\author{David Berenstein\\
\em\small
Department of Physics, University of California at Santa Barbara, CA 93106, USA}



 \maketitle

\begin{abstract}

Recent progress towards understanding a strong coupling expansion for various superconformal field theories in four dimensions is described. First, the case of the maximally supersymmetric Yang Mills theory is analyzed, as well as many calculations that can be done directly at strong coupling and matched to the AdS dual geometry. Also, this understanding is extended to other AdS duals where
the sphere is replaced by a Sasaki-Einstein manifold. Particular emphasis is made on matching exactly part of the supergravity  dual spectrum of various of these field theories by using wave function methods.

{\em Contribution to the proceedings of the Osaka ``Progress of String theory and quantum Field theory" Workshop, OCU, Dec. 2007}
\end{abstract}

\section{ Introduction}

It has been ten years since the AdS/CFT correspondence was formulated \cite{Malda}. In its simplest incarnation it is an exact quantum duality type IIb string theory on asymptotically $AdS_5\times S^5$ geometries and the maximally supersymmetric Yang Mills theory on $S^3\times \BR$. 
A basic understanding of the symmetries shows that the the isometries of $AdS_5\times S^5$
given by the global symmetries of the field theory. It was also understood from the very beginning that the radius of curvature of the geometry in string units scales as
$R^4\sim g_{YM}^2 N$, which is the t' Hooft coupling \cite{'tH}. In order to compare the two sides of the duality, one is forced to consider either a regime in gravity where $R$ is small and string corrections are large, but the field theory is perturbative, or one has to consider a regime where $R$ is large and therefore the field theory is strongly coupled. If moreover one requires that the string coupling constant is small, then $N$ is large. 

As such, the AdS/CFT correspondence is a strong-weak coupling duality for the 't Hooft coupling. This conjecture generalizes to many other families of four dimensional superconformal field theories.  The objective of this paper is to explore recent progress on the understanding of the ${\cal N}= 4 $ SYM at strong coupling and how this understanding also lets us solve to a large extent other conformal field theories in four dimensions at strong coupling. These other conformal field theories in gravity result from replacing the five-sphere by some Sasaki-Einstein manifold \cite{KWHor}. The gist of the argument is that the dominant dynamics for all of these cases is a local ${\cal N}=4 $ SYM theory in the regions of field space that dominate the dynamics \cite{B,BH}.
The understanding of these special regions in field space is done by a self-consistent approximation which is essentially non-perturbative.
This proposal for strong coupling makes many geometrical aspects of the higher dimensional gravity theory manifest. The expansion also gives a collection of tools that permit one to do various dynamical calculations directly at strong coupling. This permits one to produce a lot of evidence justifying the proposed strong coupling expansion. 

The paper will explore in succession the approximation of strong coupling to ${\cal N}=4 $ SYM, making emphasis on how degrees of freedom separate into eigenvalues and off-diagonal degrees of freedom at strong coupling. This division can then be used to show how strings emerge and how one would compute their energies. For other field theories it will then be shown that this same type of division of degrees of freedom persists, but where the diagonal degrees of freedom are related to solutions of the moduli space of vacua of the theory.  I will also explain briefly how one recovers part of the non-BPS dual supergravity spectrum on $AdS_5\times X$ from this proposal for strong coupling dynamics, as well as the future prospects for this approach to understand the AdS/CFT duality.

\section{Commuting matrix quantum mechanics and eigenvalue gases}

In order to understand what happens to ${\cal N}=4 $ SYM at strong coupling, it is useful to start with the study of supersymmetric states. These have a chance at being protected from weak to strong coupling, not just in how one counts them, but perhaps also to some extent their dynamics will be simple. By using the operator-state correspondence associated to four dimensional field theories on a sphere times time, one finds that the BPS states that respect one eight of the supersymmetries are in one to one correspondence with chiral ring operators in the field theory. At weak coupling one can further use the operator state correspondence to show that these supersymmetric states depend only on the constant modes of the scalars on the sphere \cite{Bdrop,BlargeN} (plus some fermion polarizations that we will ignore because fermions do not get classical vacuum expectation values).

This suggest that in order to study supersymmetric states it suffices to study the quantum mechanics of the constant modes on the sphere at weak coupling. It seems reasonable to assume that this will also hold classically at strong coupling and that supersymmetry will make semiclassical arguments exact. The effective dynamics of such modes is similar to the dimensional reduction on the sphere. The effective dynamics is given by a truncated action

\begin{equation}
S\sim \int \tr \left\{\sum_i \frac 12( D_t X^i)^2-\frac 12 (X^i) ^2 +\sum_{ij}  \frac 1{8\pi^2} g_{YM}^2 [X_i,X_j]^2\right\}
\end{equation}
where we need to keep the constant mode of the gauge field on the sphere in order to impose the gauge constraints on the configurations. The effective mass term arises from the conformal coupling of the scalars to the curvature of $S^3\times \BR$. Here we have also have rescaled the field so that the kinetic term is canonical. The factor of $8\pi^2$ reflects the volume of the $S^3$.

To analyze this system at strong coupling, we can use a naive large N counting where $X$ is of order $\sqrt N$ (the typical size of an eigenvalue of a random matrix). We find that the kinetic and quadratic terms are of order $N^2$, while the commutator term is of order $\lambda N^2$, where $\lambda$ is the 't Hooft coupling constant. This suggests that at strong coupling the quartic potential dominates the dynamics, and therefore it should be minimized. The minimization of the commutator potential tells us that the low energy dynamics is dominated by configurations of commuting matrices. These are quantized with the quadratic pieces of the action (one can think of this as treating the kinetic term and mass term as a perturbation on the configurations of zero potential).

This produces a model of gauged commuting matrix quantum mechanics. One should remember that commuting matrices describe the moduli space of vacua of the field theory (they solve the D and F-term constraints). This is important for generalizations to other quantum field theories.

The matrices $X$ are hermitian and they commute, so they can be diagonalized simultaneously by a gauge transformation.
The only relevant dynamical variables are the eigenvalues of the matrices themselves. If we make a diagonal ansatz in the classical action, we find that the system appear to be free. However, when one computes the effective Hamiltonian, one needs to take into account that we have changed variables from general commuting matrices to diagonal matrices, so that the effective Hamiltonian does not have a standard Laplacian. Instead it is modified by a measure term, which is the volume of the gauge orbit of a configuration \cite{BlargeN}. The full answer is given by
\begin{equation}
H_{eff} = \sum_i -\frac 1{2\mu^2} \nabla_i \mu^2 \nabla_i +\frac 12 \vec x_i^2
\end{equation}
where we have a sum over eigenvalues. These are collected into a six vector $\vec x$ (we have six matrices), and the eigenvalues are labeled by $i$ (there are $N$ eigenvalues). The measure is given by a generalized Van der Monde determinant
\begin{equation}
\mu^2 = \prod_{i<j} |\vec x_i -\vec x_j|^2
\end{equation}
It turns out that the Gaussian wave function is an eigenfunction of this Hamiltonian,
\begin{equation}
\psi_0 \sim \exp\left( -\sum_i \frac{\vec x_i^2}2\right)
\end{equation}
and that the probability measure on eigenvalue space (given by the square of the wave function times the measure) can be interpreted as a Boltzman gas of $N$ particles in six dimensions with a confining quadratic potential and repulsive logarithmic interactions
\begin{equation}
p (\vec x) \sim \exp( -\sum_i \vec x_i^2+  \sum_{i<j}\log(|\vec x_i-\vec x_j|^2)
\end{equation}
In the thermodynamic limit (large N) one can solve the Boltzmann gas by a saddle point approximation solving for the density of eigenvalues. One finds that the distribution of eigenvalues is singular, concentrated on a sphere \cite{BlargeN,BCV}
\begin{equation}
\rho = N\frac{\delta(|\vec x|-r_0)}{r_0^5 Vol(S^5)}
\end{equation} 
and $r_0= \sqrt {(N/2)}$. One can guess this form by the $SO(6)$ R-symmetry of the system.
This serves as a consistency check of our analysis, where we said that the matrices where of order $\sqrt N$. The size could have been renormalized at strong coupling by a function of $\lambda$. This can also be analyzed numerically by Monte Carlo methods to see how well the eigenvalue distribution is captured by the saddle point calculation \cite{BCott}.

The end result is that the eigenvalues form a five-sphere. This sphere should be identified with the $S^5$ of the dual gravity theory and suggests an origin for the higher dimensional geometry of the dual theory. Since the geometry is encoded non-trivially in the quantum wave function of the system, this is a setup where one can talk about emergent geometry in a precise sense. 

\section{Off-diagonal degrees of freedom and strings}

So far, the discussion has centered on the eigenvalues, and we have ignored all other degrees of freedom. The analysis has relied on working in a low energy expansion. Off-diagonal modes should be heavy for this to make sense. One can work with these off-diagonal modes perturbatively. One can check that to quadratic order the off-diagonal modes 
have masses of order $m^2\sim \lambda$ which is taken to be large in order to expand at strong coupling. 

Being more careful, one notices that a typical off-diagonal mode connects two different eigenvalues $\vec x^i$ and $\vec x^j$, and one can see that the mass of the off-diagonal modes is given by
\begin{equation}
m^2_{ij} = 1+ g^2 |\vec x_i-\vec x_j|^2
\end{equation}
where $g$ differs from $g_{YM}$ by normalization factors. At strong coupling the one can be dropped, and one finds that the mass is essentially given by the Euclidean distance between the two particles on $\BR^6$. Here we interpret eigenvalues as D-particles in the usual way. One can also imagine drawing an off-diagonal mode as a line segment connecting two diagonal degrees of freedom.
We see this way that the distance on $\BR^6$ matters to determine the off-diagonal mode structure, and that the metric information in $\BR^6$ is physical. 

The typical diagonal configuration breaks the gauge group to $U(1)^N$, and the off-diagonal modes are charged. Gauge invariance requires that for any off-diagonal perturbation entering into a D-particle, there is a corresponding off-diagonal perturbation exiting from it. This requires that the off-diagonal excitations form closed polygons, and this suggests that the off-diagonal modes give rise to a structure of excitations that is realized by some form of closed string (at this stage it is a combinatorial approximation to smooth curves on $S^5$).

Indeed, one can use this understanding to show that the energies of strings are reproduced exactly\cite{BCV} in the plane wave limit  and also the so-called giant magnon configurations \cite{BMN} (see also \cite{Beis}). This shows that the five-sphere has the right size in string units, as one is reproducing the energies of many non-trivial string states. Indeed, the field theory calculation, which is done in a saddle point approximation, suggested that the worldsheet momentum on the quantum field theory spin chain model was given by an angle on the sphere. This was verified later in the semiclassical calculation of giant magnons by Hofman and Maldacena \cite{HM}. One can also study bound states of giant magnons \cite{Dorey} and the strong coupling calculations have a match as well \cite{BV}, but this is less well understood since the quantum numbers of the corresponding bound states are different.

\section{Other $N=1$ superconformal field theories}

Now that the structure of ${\cal N}=4 $ SYM has been understood at strong coupling, it is important to see if there is a way to extend the techniques developed in the well known case to other superconformal field theories that are less well understood. In particular, for many conformal field theories one can check explicitly that there is no weak coupling limit, as the fundamental fields acquire large anomalous dimensions \cite{LS}. This makes any treatment of such theories difficult in general. Some of these issues were overcome in \cite{B}. It was argued that there might be some regimes (regions of field space) in the field theory where a classical calculation might be valid, even though the field theory is
in general always strongly coupled. This assumption is in some sense required in order to be able to say anything useful for such a field theory. One can begin with a few assumptions regarding this classical regime. These are that the field theory is well described by a classical supersymmetric lagrangian with a (non-trivial) Kahler potential, a superpotential and  super Yang Mills sectors. One can them impose on this classical lagrangian the conditions that lead to a superconformal classical theory.

The assumptions are that each chiral field has an $R$-charge, and that this is a holormorphic weight ( which is equal to it's conformal weight). For the kinetic term of the scalar to be invariant under local rescalings of the background metric, we find that the Kahler potential has to be a scaling function of weight $(1,1)$ with respect to the holomorphic and anti-holomorphic degrees of freedom. Also, one finds a non-minimal coupling of the scalars to the background curvature. This is almost identical to the case of ${\cal N}=4 $ SYM. 

One can also prove that the supersymmetric classical states on a sphere that are dual to the chiral ring are simple. They have no gauge field excitations, only constant modes on the sphere are excited and the $F$ and $D$-terms must vanish. This set of conditions states that up to gauge equivalence they are in one to one correspondence with the moduli space of vacua of the field theory. This can be analyzed in detail as it depends only on the superpotential of the theory.

The upshot of this calculation is that configurations of commuting matrices in ${\cal N}=4 $ SYM are replaced by configurations on the moduli space of vacua of the theory. These moduli spaces are well understood in some cases, and they are for the most part given by a symmetric product
\begin{equation}
{\cal M} = Sym^N V
\end{equation}
where $V$ is a Calabi-Yau cone (these are the situations where the dual gravity theory has  Freund-Rubin ansatz \cite{FR}, and the $S^5$ has to be replaced by a Sasaki Einstein manifold $X$, so that $V$ is the cone over $X$). In essence, this is what the D-brane moduli space for D-branes near a Calabi-Yau cone singularity should look like. Moreover, it was argued that the metric on $V$ should be Ricci flat.

In essence, instead of having a gas of D-particles on $\BR^6$, one ends up with a gas of D-particles on $V$. This dynamics follows the axions of D-geometry as described by Douglas, Kato and Ooguri \cite{DKO}. Basically, away from the singularity at the origin, one has an effective dynamics that is locally equivalent (on $V$) to ${\cal N}=4 $ SYM, so that the previous insights can be applied. This is, one finds a window where one can do effective field theory as one does in ${\cal N}=4$ SYM, but where the global structure of the moduli space of vacua is modified.

The most important effect is that ``off-diagonal" energies are proportional to the distance between D-particles on $V$, for any pair of particles that is near each other, and that this is true for all polarizations. This is how the metric on $V$ becomes important.

To complete the effective dynamics one also needs an effective Hamiltonian on the moduli space of vacua. As in the case of ${\cal N}=4 $ SYM, the system is originally written in terms of a gauge theory (or collection of general matrices), whereas the notion of D-particles as eigenvalues also requires one to compute some analogue of the volume of the gauge orbit. However, if one tries to understand how this dynamics could be independent of the choice of presentation of the field theory under Seiberg dualities \cite{S}, one finds that a naive counting of the volume of the gauge orbit is not the correct answer.
A proposal for this measure term was provided in joint work with Hartnoll \cite{BH}. This proposal is as follows. 

First, the measure factorizes, so that
\begin{equation}
\mu^2 = \exp(-\sum_{i<j} s_{ij})
\end{equation}
and that the functions $s$ satisfy a differential equation on moduli space
\begin{equation}
-\nabla^6 s(x,x') = 64\pi^3 \delta^{(6)}(x,x')
\end{equation}
This was guessed by noticing that in ${\cal N}=4 $ SYM the logarithm in the measure satisfies these properties. Moreover, it was found that in the case of orbifolds, the method of images would give a function that satisfies the same constraints \cite{BC1}.

From here, one gets an effective Hamiltonian
\begin{equation}
H= \sum_{i}\left[-\frac 1{2\mu^2}\nabla_i (\mu^2\nabla_i) +K_i \right]\label{eq:Hamse}
\end{equation}
where $K_i$ is the Kahler potential of $V$,  and surprisingly, one can also solve the ground state wave function
\begin{equation}
\psi_0= \exp(-\sum_i K_i)\label{eq:wfse}
\end{equation}
Again, one can use the probability density associated to this wave function  to determine the important physical configurations. One has a gas of particles on $V$ subject to two body logarithmic repulsions and to an external confining force (essentially $K$ acts as a harmonic oscillator potential).

One can prove \cite{BH} that in the saddle point limit all the D-particles are at the same distance (of order $\sqrt N$) from the tip of the cone, and that they form a uniform distribution. This information basically states that in the geometry one has replaced the five sphere with a Sasaki-Einstein metric.
It is also possible to show that one reproduces exactly a particular plane wave limit spectrum for all $X$, where the plane wave limit is taken along an $R$-charge isometry and that this determination is reliable at a strong effective t'Hooft coupling. This requires also the introduction of short off-diagonal modes. 

The upshot is a unified dynamics for all of these field theories that singles out ${\cal N}=4 $ SYM and explains why if ${\cal N}=4 $ SYM can be described at strong coupling by type IIB string theory, then this is true also for a host of different superconformal field theories (exactly those whose moduli space has the requisite symmetric product structure).

\section{Supergravity spectrum}

It is instructive to explain how some of the calculations that show the match at strong coupling are done. 
In particular, a single graviton propagating on $AdS_5\times X$ should be considered as a low lying excitation of the system. However, most gravitons will also have angular momentum on $AdS_5$, and they are not captured by the truncation to the low lying degrees of freedom we have performed, that has all degrees of freedom constant on the $S^3$ and therefore they do not carry angular momentum. Fortunately, various such gravitons do not carry angular momentum either, and it is possible to compare some of these with particular excitations of the low energy effective system.

With the understanding that in the effective field theory we have a finite number of degrees of freedom, 
with a Hamiltonian given by equation (\ref{eq:Hamse}) and that we have solved the system in terms of wave function of these degrees of freedom (\ref{eq:wfse}), to find excitations it would be required to have a wave function describing these excitations.

Also, in the operator language, chiral operators can be described in terms of  traces of products of holomorphic 
fields, where the large $N$ counting suggests that each traces is interpreted as a single graviton. This suggests a form for the appropriate wave function of a supergravity mode
\begin{equation}
\psi\sim \tr(f(x)) \psi_0 \label{eq:wfexc}
\end{equation}
where $f$ is now a function on $V$. It is also known that in the case of ${\cal N}=4 $ SYM, the $f$ that show up are symmetric traceless combinations of the fields, and all of these turn out to be harmonic functions on $\BR^6$. The ansatz for a wave function that we make is that $f$ is a harmonic function on $V$. Then we can replace this ansatz in the Hamiltonian \ref{eq:Hamse} and verify if it is an eigenfunction or not. We should check this ignoring $1/N$ corrections.

The metric on $V$ is of the form 
\begin{equation}
ds^2 = dr^2 +r^2 dX_5^2
\end{equation}
where $X^5$ is the Sasaki-Einstein metric. This allows for a separation of variables in $r$ and $X$.
One finds that harmonic functions on $V$ are related to general eigenvalues of the laplacian on $X$. Moreover, one finds that as functions of $r$, they are scaling functions where $h(r) = r^\lambda$, and $\lambda$ is given by 
\begin{equation}
\lambda= -2+\sqrt{4+\nu^2}
\end{equation}
where $\nu$ is the eigenvalue of the Laplacian of $X$. One also finds that the wave function (\ref{eq:wfexc}) is an eigenfunction of the Hamiltonian to a very good approximation, with energy $\lambda$ over the ground state. This depends on using properties of the saddle point distribution of particles and assuming that the modified wave function does not alter this distribution too much. This is a good approximation on the thermodynamic limit $N\to \infty$.
Thus the possible values of $\lambda$ describe the different dimensions of operators that are allowed at strong coupling. Notice that we are departing from the traditional idea that operators at strong coupling are simply polynomials in the fundamental fields. Instead, the characteristics of $f$ determine the spectrum of dimensions of operators in the conformal field theory.

One finds a similar formula for the energies of the gravitational fluctuations of $AdS_5\times X$ that mix the five form field strength and the metric and the spectrum of the Laplacian on $X$ appears in the same way \cite{Gub}. Indeed, one can match exactly half of those modes to the wave functions that we have described so far. The harmonic functions that are holomorphic are special, and those happen to be polynomials in the variables describing the algebraic geometry of $V$.

\section{Outlook}

So far, in this review I have described how there is a natural notion of emergent geometry that appears at strong coupling for a large family of superconformal field theories. This notion of geometry provides a simple notion of locality, which is based on the idea that the energies of off-diagonal degrees of freedom connecting 
points on the geometry are proportional to the distances in the geometry, so faraway points are decoupled.  It was also described that in such a geometry there is a universal dynamics: it is given by a local ${\cal N}=4 $ SYM on the moduli space of vacua for the relevant configurations. It has also been shown that a lot of calculations that were first understood in gravity can now be performed at strong coupling in field theory and that one finds an exact match for a large class of observables.

However, a complete understanding of the AdS/CFT duality is still beyond reach. Although one has been able to make good progress on the understanding of strongly coupled phenomena (at least as it pertains to the systems studied here), there is still a lot to be done. The radial direction of AdS is understood very poorly, as well as the reason as to why time is warped in the gravitational theory. Although it is possible to get this information from the isometry group of $AdS$, this is a kinematical rather than a dynamical reason. This means that when one breaks conformal invariance one does not have an understanding of how this would affect the geometry. There is also a need to understand how heavy objects produce a gravitational interactions. This is something that has not been done so far, although there seems to be no a-priori obstacle to doing these calculations.
One should be optimistic about the general prospects for the future understanding of these issues. The techniques that have been developed allow for more involved calculations and the fact that there is some understanding of locality in higher dimensions can help a lot towards developing intuition and making setups that have some direct correspondence with the gravity theory.

One should also be content with the fact that since  the developments so far have used very nice looking wave functions, it is not too hard to make simulations of their associated probability densities where direct analytical calculations might be very hard.

\section*{Acknowledgements}
The author would like to thank D. Correa, R. Cotta, S. Hartnoll, R. Leonardi, S. Vazquez for various collaborations related to this paper. Work supported in part by 
 the U.S. Department of Energy, under grant DE-FG02-91ER40618.

\end{document}